\documentclass{aastex63}

\usepackage[version=3]{mhchem}
\usepackage{booktabs}
\usepackage{color,soul}
\usepackage{savesym}
\savesymbol{tablenum}
\usepackage{siunitx}
\usepackage[left]{lineno}
\restoresymbol{SIX}{tablenum}
\DeclareSIUnit{\molecule}{molecule}
\definecolor{mypink}{rgb}{0.858, 0.188, 0.478}
\definecolor{mypink3}{cmyk}{0, 0.8808, 0.9429, 0.3412}
\definecolor{mygreen}{cmyk}{0.59, 0.00, 0.99, 0.10}
\definecolor{lightblue}{rgb}{0.659,0.8706,1}
\definecolor{blu}{rgb}{0.2039,0.388,1}
\sethlcolor{lightblue} \setstcolor{red}

\renewcommand{\thefootnote}{\fnsymbol{footnote}}
\renewcommand\arraystretch{1.2}
\submitjournal{ApJ}

\shorttitle{formation of phosphorus monoxide}
\shortauthors{García de la Concepci\'on et al.}
\graphicspath{{./}{figures/}}
\begin{document}

\title{Formation of phosphorus monoxide (PO) in the interstellar medium: insights from quantum-chemical and kinetic calculations}

\correspondingauthor{Juan García de la Concepci\'on}
\email{jgarcia@cab.inta-csic.es}
\correspondingauthor{Vincenzo Barone}
\email{vincenzo.barone@sns.it}
\correspondingauthor{Cristina Puzzarini}
\email{cristina.puzzarini@unibo.it}

%PYou can also add your ORCID
%\author[orcid-author1]{Author 1}
\author[0000-0001-6484-9546]{Juan García de la Concepción}
\affiliation{Centro de Astrobiolog\'{\i}a (CSIC-INTA), Ctra. de Ajalvir Km. 4, Torrej\'on de Ardoz, 28850 Madrid, Spain}

\author[0000-0002-2395-8532]{Cristina Puzzarini}
\affiliation{Department of Chemistry ``Giacomo Ciamician'', University of Bologna,
Via F. Selmi 2,
Bologna, 40126, Italy}

\author[0000-0001-6420-4107]{Vincenzo Barone}
\affiliation{Scuola Normale Superiore,
Piazza dei Cavalieri 7,
Pisa, 56126, Italy}

\author[0000-0003-4493-8714]{Izaskun Jim\'enez-Serra}
\affiliation{Centro de Astrobiolog\'{\i}a (CSIC-INTA), Ctra. de Ajalvir Km. 4, Torrej\'on de Ardoz, 28850 Madrid, Spain}

\author[0000-0002-8871-4846]{Octavio Roncero}
\affiliation{Instituto de F\'{\i}sica Fundamental (IFF), C.S.I.C., Serrano 123, 28006 Madrid, Spain}



\begin{abstract}
In recent years, phosphorus monoxide (PO) --an important molecule for prebiotic chemistry-- has been detected in star-forming regions and in the comet 67P/Churyumov-Gerasimenko.  These studies have revealed that, in the interstellar medium, PO is systematically the most abundant P-bearing species, with abundances that are $\sim$1-3 times greater than those derived for phosphorus nitride (PN), the second most abundant P-containing molecule. The reason why PO is more abundant than PN remains still unclear. Experimental studies with phosphorus in the gas phase are not available, probably because of the difficulties in dealing with its compounds. Therefore, the reactivity of atomic phosphorus needs to be investigated using reliable computational tools. To this end, state-of-the-art quantum-chemical computations have been employed to evaluate accurate reaction rates and branching ratios for the P + OH $\rightarrow$ PO + H and P + H$_2$O $\rightarrow$ PO + H$_2$ reactions in the framework of a master equation approach based on ab-initio transition state theory. The hypothesis that OH and H${_2}$O can be potential oxidizing agents of atomic phosphorus is based on the ubiquitous presence of H${_2}$O in the ISM. Its destruction then produces OH, which is another very abundant species. While the reaction of atomic phosphorus in its gound state with water is not a relevant source of PO because of emerged energy barriers, the P + OH reaction represents an important formation route of PO in the interstellar medium. Our kinetic results show that this reaction follow an Arrhenius behavior, and thus its rate coefficients (\(\alpha\)=2.28$\times$10$^{-10}$ cm${^3}$ molecule$^{-1}$ s$^{-1}$, \(\beta\)=0.16 and \(\gamma\)=0.37 K) increase by increasing the temperature.
\end{abstract}



\section{Introduction} 
\label{sec:intro}

Phosphorus (P) is a key element for life, indeed being present in all living systems \citep{Gulick57,Macia05,Fernandez-Garcia,Schwartz2006}. It is an essential constituent of biomolecules, where it plays a key role in crucial processes such as information transfer and replication (RNA and DNA), formation of cellular membranes (phospholipids), and energy production in living cells (ATP). 

Compared to living systems \citep[P/H = 10$^{-3}$;][]{Fagerbakke96}, the abundance of P (with respect to hydrogen) in the Universe is low \citep[P/H = 2.8$\times$10$^{-7}$;][]{Grevesse98}.
P is known to form in massive stars and it is distributed throughout the interstellar medium (ISM) by supernova explosions \citep{Koo2013}. It is believed that P became available in the early Earth during  the Late Heavy Bombardment period, i.e. between 4.1 to 3.8 billion years ago, when this element was delivered by the impact of small bodies such as meteorites \citep{Pasek05} and comets similar to Halley and 67P/Churyumov-Gerasimenko (67P/C-G) \citep[hereafter 67P/C–G;][]{Altwegg16}. 

Phosphates PO$_{4}^{2-}$, and their derivatives, are the source of P in biomolecules. Thus, from a chemical point of view, the oxidized forms of P are of particular interest. Among oxygenated P-bearing species, phosphorus monoxide (PO) is the simplest molecule, and it can be seen as the building block of small biomolecules \citep{Douglas20}.
PO has already been detected in different circumstellar environments and star-forming regions, where several observational studies have systematically pointed out that PO is more abundant than PN (the second most abundant P-bearing species detected in molecular clouds) by a factor of $\sim$1-3 \citep[][]{Lefloch,bergner19,victor16,victor2018,Victor2020,bernal21}. Recently, \citet{Victor2020} have shown that PO, and not atomic phosphorus, is the reservoir of P in comet 67P/C-G. 

Several mechanisms have been proposed to explain the production of P-bearing species in the ISM such as shock-induced formation \citep{Yamaguchi11,Aota,Lefloch,Mininni18,victor2018}, gas-phase reactions in the cold collapse phase \citep{victor16} and gas-phase processes at high temperatures in massive hot cores \citep{charnley94}. However, the question why  PO is more abundant than PN in the gas phase is still open \citep{bergner19,bernal21}.
\citet{Izaskun18} revisited the chemistry of phosphorus in the ISM under different conditions and accounting for energetic phenomena (protostellar heating, cosmic rays, UV-photon radiation, and shock waves) in order to derive the PO/PN abundance ratio and understand how this is affected by different environments. To reproduce the PO/PN abundance ratios obtained in the molecular outflows of shocked regions \citep[e.g.][]{Yamaguchi11, Lefloch}, the gas-phase reaction between P and the OH radical was incorporated in the chemical network \citep{Izaskun18}. However, the lack of kinetic information for such a reaction compelled the authors to assume that the rate constant is equal to that of the association reaction between N and OH yielding NO + H \citep[see Table$\,$4 in][]{Izaskun18}. Indeed, N + OH is the main formation route of NO in star-forming regions and it is able to explain the NO abundances measured in protostellar envelopes and shocks \citep{Codella18}.

At variance with N, the reactivity of atomic phosphorus with O-bearing species such as OH and water, but also atomic O, has so far only marginally been studied. The radiative association between P and O to produce PO in its doublet ground state, has been investigated by \citet{andreazza16}. Their quantum-chemical calculations show that this reaction is inefficient in the ISM because the derived rate constants are too small; indeed, in the range between 300 K and 14000 K, they vary from 1.6$\times$10$^{-24}$ to 2.0$\times$10$^{-18}$~cm$^{3}$~molecule$^{-1}$~s$^{-1}$ \citep{andreazza16}. The reaction of atomic phosphorus (${^4}S$ and ${^2}D$) with molecular oxygen, O$_2$, has been studied from both experimental and theoretical points of view \citep{Douglas19}. However, several discrepancies have been found in the kinetics of the reaction between P in its ground state (${^4}S$) and O$_2$, which seems to show an unusual temperature dependence \citep{Douglas19,Henshaw87,Husain77,Husain78,Clyne82}. Finally, to the best of our knowledge, the gas-phase P + OH $\rightarrow$ PO + H and P + H$_2$O $\rightarrow$ PO + H$_2$ reactions have not yet been investigated neither experimentally nor theoretically.

In this paper, we report the main results of a comprehensive computational study on the formation of phosphorus monoxide ($^2\Pi$) from the reactions of atomic phosphorus (P) with the hydroxyl radical (OH) and with water (H$_2$O).  
Although atomic phosphorus in the ISM should be in the electronic ground state ${^4}S$, we have also considered its metastable (first) excited state ${^2}D$ because its formation is associated with high energetic processes \citep{Koo2013}. As potential oxidizing agents of atomic phosphorus, H${_2}$O and OH have been employed. The former species is highly abundant in the ISM and is destroyed through different mechanisms to yield OH (\citet{Viti11,tappe08}). 
The paper is organized as follows. Section$\,$\ref{sec:compdet} describes the computational methodology. Section$\,$\ref{sec:results} presents the results on the potential energy profiles and kinetics for both the reactions considered. In Section$\,$\ref{sec:discussion} we discuss the astrophysical implications and, finally, in Section$\,$\ref{sec:conclusions} we summarize our conclusions.

\section{Computational methodology} 
\label{sec:compdet}

The starting point for the study of the formation pathways of ${^2}$PO is the identification of the potential reactants and the analysis of the corresponding reactive potential energy surface (PES) to accurately characterize all stationary points from both a structural and energetic point of view. In a second step, kinetic calculations are performed under the very low temperatures (10-100 K) and pressures (10-10$^7$ cm$^{-3}$ in terms of number density) typical of the ISM.

\subsection{Reactive potential energy surface}
\label{pes:det}

To characterize the reactive PES, we have followed a well-established computational strategy \citep{Vazart2016,lupi:h2s,staa1652} that has recently been employed in the investigation of the formation mechanisms of different interstellar molecules \citep{Tonolo2020,Molecules,staa1652,vincenzo21,5.0038072}. This strategy consists of the following steps.

First, all stationary points of the reactive PES have been located and characterized using the double-hybrid rev-DSD-PBEP86 functional \citep{kozuch11,santra19} combined with the D3(BJ) corrections to incorporate dispersion effects \citep{grimme10,grimme11}. This functional has been used in conjunction with the jun-cc-pVTZ basis set, thereby using the set with an additional $d$ function on P, jun-cc-pV(T+d)Z \citep{papajak,dunningJr.01,prascher11}.
In the following, this level of computation will be simply referred to as revDSD.
The nature of the stationary points located on the PES has been confirmed by Hessian evaluations at the same level of theory, which also provide the corresponding harmonic zero-point vibrational energy (ZPE) corrections. To ensure the right connection between transition states and minima, intrinsic reaction coordinate (IRC) analyzes have been performed throughout the PESs characterization \citep{1.1927521} To further check the reliability of the revDSD structural characterizations, for the reaction involving the OH radical, the stationary points of the reactive PES have also been optimized using the explicitly-correlated CCSD(T)-F12b method \citep{ref:f12,werner2011explicitly,knizia2009simplified} in conjunction with the cc-pVTZ-F12 basis set \citep{ref:vnzf12}, within the frozen-core (fc) approximation. Furthermore, we compared our results with those available in the literature for the HOP and HPO species \citep{FRANCISCO2003303,PUZZARINI2006238}. In passing we note that revDSD geometries well agree with those obtained using different coupled-cluster formulations.

In a second step, single-point energy calculations, on top of revDSD geometries, have been performed by means of the composite scheme denoted as HEAT-like because based on the HEAT protocol \citep{heat,heat2,heat3}. As detailed in \citet{D0CP00561D,lupi:h2s,staa1652}, the HEAT-like scheme relies on the additivity approximation, with the different contributions required for obtaining highly-accurate results evaluated at the best possible level and then combined together:

\begin{equation}
E_{tot} =  E^{\infty}_{\mathrm{HF-SCF}} + \Delta E^{\infty}_{\mathrm{CCSD(T)}} + \Delta E_{\mathrm{CV}} + \Delta E_{\mathrm{fT}} +  \Delta E_{\mathrm{fQ}} + \Delta E_{\mathrm{REL}} + \Delta E_{\mathrm{DBOC}}  \; .
\label{heat}	   
\end{equation}
\vspace{0.5mm}

In the expression above, $E^{\infty}_{\mathrm{HF-SCF}}$ and $\Delta E^{\infty}_{\mathrm{CCSD(T)}}$ denote the extrapolation to the complete basis set (CBS) limit of the HF-SCF energy using the the exponential formula by \citet{Feller1993} and the CCSD(T) correlation energy (within the fc approximation) extrapolated to the CBS limit with the $n^{-3}$ expression \citep{Helgaker1997}, respectively. The correlation-consistent cc-pV$n$Z basis sets \cite{Dunning-JCP1989_cc-pVxZ} have been employed in conjunction with these calculations, with $n$=Q, 5 and 6 being chosen for the HF-SCF extrapolation, and $n$=Q and 5 for the CCSD(T) correlation energy. The $\Delta E_{\mathrm{CV}}$ term allows for incorporating the core-valence correlation correction, evaluated as energy difference between all-electron (ae) and fc-CCSD(T) computations in the same basis, which is --in the present case-- the cc-pCVQZ set \citep{WD95,PD02}. In a similar manner, corrections due to the full treatment of triple, $\Delta E_{\mathrm{fT}}$, and quadruple, $\Delta E_{\mathrm{fQ}}$, excitations are computed as energy differences between CCSDT \citep{ccsdt1,ccsdt2,ccsdt3} and CCSD(T) and between CCSDTQ \cite{ccsdtq} and CCSDT (all within the fc approximation) employing the cc-pVTZ and cc-pVDZ basis sets, respectively. The diagonal Born-Oppenheimer correction \citep{dboc1,dboc2,dboc3,dboc4}, $\Delta E_{\mathrm{DBOC}}$, and the scalar relativistic contribution \citep{rel1,rel2} to the energy, $\Delta E_{\mathrm{REL}}$, are also included. The former correction has been computed at the HF-SCF/aug-cc-pVTZ level \citep{KDH92}, whereas the relativistic corrections have been obtained at the ae-CCSD(T)/aug-cc-pCVTZ level including the (one-electron) Darwin and mass-velocity terms. Incorporation of CCSDTQ computations in the composite scheme should account for non-dynamical electron correlation effects; to further check this point, for the stationary points showing a multireference character (deduced from the value of the T$_1$ diagnostic; \citet{qua.560360824}), the correction due to pentuple excitations has also been considered and incorporated as energy difference between CCSDTQP \citep{ccsdtq} and CCSDTQ computations carried out within the fc  approximation and in conjunction with the cc-pVDZ basis set. 
Finally, HEAT-like energies have been augmented by anharmonic ZPE corrections evaluated at the revDSD level within vibrational perturbative theory to second order (VPT2; \citet{Barone2004}).

The conclusive test on the reliability of coupled-cluster calculations has been provided by multiconfigurational computations using the n-electron valence state perturbation theory (NEVPT2) method \citep{angeli01,angeli012}. This has been employed in conjunction with the minimally augmented ma-def2-(T+d)ZVP basis set. The active space chosen for the phosphorus atom is three electrons in 3 \textit{p} atomic orbitals (3,3).  
For the hydroxyl radical, a (5,4) active space (i. e. the two \textit{p}\(\pi\) orbitals and the pair \textit{p}\(\sigma\) and \textit{p}\(\sigma\)*) has been selected instead. The overall active space employed is (8,7). Since the NEVPT2 test for the stationary points showing relevant non-dynamical correlation effects conclusively demonstrated the reliability of our coupled-cluster approach, we will not mention further NEVPT2 results.

All DFT and VPT2 calculations have been carried out with the Gaussian software \citep{g16}, while those for the HEAT-like scheme have been performed using the CFOUR \citep{cfour,cfour-new}, except those including quadruple and quintuple excitations (CCSDTQ,CCSDTQP) which have been performed with the MRCC code \citep{mrcc} interfaced to CFOUR. NEVPT2 computations have been carried out with the ORCA software \citep{orca}, while explicitly correlated CCSD(T)-F12b calculations have been performed using the MOLPRO software \citep{molpro}.

\subsection{Kinetic calculations}
\label{Kin:Mod}

For elementary steps ruled by well-defined saddle points, the unimolecular rate coefficients were calculated using Rice-Ramsperger-Kassel-Marcus (RRKM) theory within the rigid-rotor harmonic-oscillator (RRHO) approximation \citep{ChemKin}. Barrierless bimolecular association rate constants were instead obtained employing the phase space theory (PST; \citet{pechukas1965detailed,chesnavich1986multiple}), where the attractive potential between the two fragments at large distances is described by a \(V_{MEP} = \frac{-C}{R^6}\) functional form with the $C$ constant derived from a fit of the revDSD energies. The $C$ values obtained for the barrierless processes are gathered in Table~\ref{tab6}. Semiclassical one-dimensional tunneling corrections were evaluated using the Eckart model \citep{eckart1930penetration}. Subsequently, the temperature and pressure-dependent phenomenological rate coefficients have been calculated by using a master equation approach based on transition state theory (AITSTME), employing the MESS software as master equation solver \citep{georgievskii2013reformulation}, which is available at https://github.com/PACChem/MESS. Te input information for MESS can be be provided upon request.

The global rate coefficients have been computed in the 30-400 K range, and to describe its temperature dependence, an Arrhenius-modified expression, the Arrhenius-Kooij formula \citep{kooij}, has been used: 
\begin{equation}
k(T) = \alpha \left(\frac{T}{300}\right)^\beta exp\left(-\frac{\gamma}{T}\right)
\label{AK}
\end{equation}
where \(\alpha\), \(\beta\) and \(\gamma\) are fitting parameters, determined using the computed rate coefficients at different temperatures.

\section{Results} 
\label{sec:results}

In the following, the reactive PESs of the P + OH and P + H$_2$O reactions and their thermochemistry are described in detail, then the kinetics of both reactions is addressed.

\subsection{Reaction of P(${^4}S$) and P(${^2}P$) with hydroxyl radical (OH, ${^2\Pi}$)}

\begin{figure}[t!]
  \includegraphics[width=1.1\textwidth]{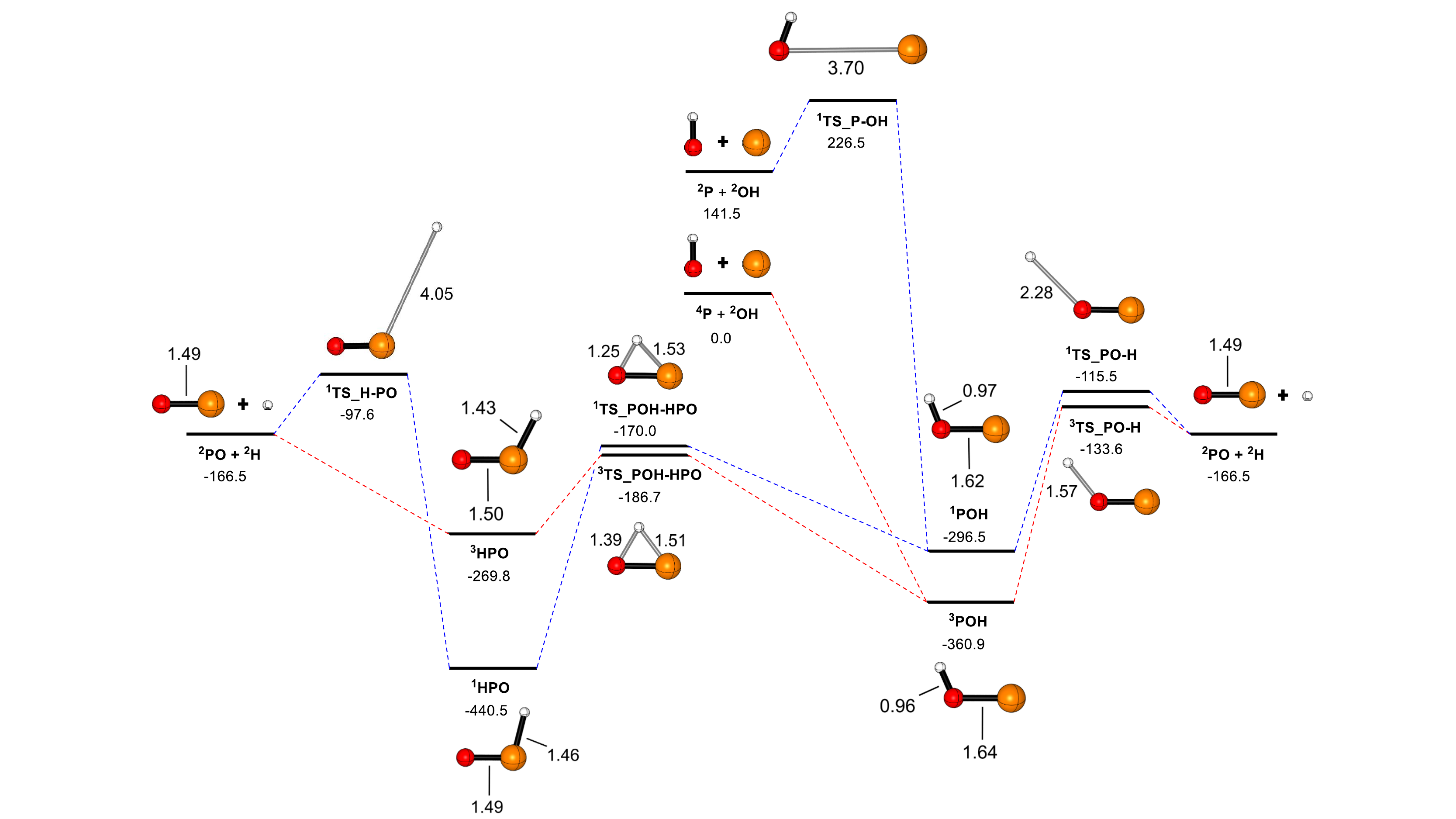}
  \caption{Singlet (blue profile) and triplet (red profile) potential energy surfaces (ZPE-corrected energies) for the formation of PO from OH + P.}
  \label{fig1}
\end{figure}

Figure~\ref{fig1} shows the HEAT-like potential energy profiles (with anharmonic ZPE corrections) for the proposed mechanisms of the reaction of atomic phosphorus with the hydroxyl radical. There are two entrance channels, one for each multiplicity, which lead to the pre-reactive complexes ${^1}$POH and ${^3}$POH. In the triplet PES, the formation of ${^3}$POH is barrierless, whereas the formation of ${^1}$POH proceeds through a transition state, ${^1}$TS$\_$P-OH. 
Therefore, only the formation of ${^3}$POH is a viable option. Indeed, to form ${^1}$POH, the system has to overcome an energy barrier of 85.0 kJ/mol. In the ISM, because of the very low temperature, only reactions with submerged barriers can occur.

\begin{deluxetable}{ccccl}
\label{tab1}
%\tablenum{1}
\tablecaption{Relative energies (in kJ/mol) for the P(${^4}S$)/P(${^2}P$) + OH(${^2}\Pi$) reaction.}
\tablewidth{0pt}
\tablehead{
%\colhead{Stationary points} & \multicolumn{2}{c}{Energy (kJ/mol)} & \colhead{Energy + ZPEanh corrected (kJ/mol)}}\\
 & revDSD${^a}$ & HEAT-like${^b}$ & \multicolumn{2}{c}{HEAT-like + ZPE\textsubscript{anh}${^c}$}}
\startdata
\hline
Triplet PES\\
\cline{1-1}
${^4}$P + ${^2}$OH &    0.0 &    0.0 &    0.0 \\
${^3}$POH          & -364.0 & -371.5 & -360.9\\
${^3}$TS\_PO-H     & -105.3 & -120.0 & -133.6\\
${^3}$TS\_POH-HPO  & -166.4 & -182.5 & -186.7\\
${^3}$HPO          & -262.1 & -272.5 & -269.8\\
${^2}$PO + ${^2}$H & -144.8 & -151.5 & -166.5\\
${^1}$PH + O(${^3}P$) & 267.3 & --- & ---\\
\hline 
Singlet PES ${^d}$\\
\cline{1-1}
${^2}$P + ${^2}$OH &  175.4 &  141.5 &  141.5\\
${^1}$TS\_P-OH     &  209.0 &  225.7 &  226.5 &(225.8) \\
${^1}$POH          & -287.7 & -308.0 & -296.5\\
${^1}$TS\_PO-H     &  -57.5 & -102.7 & -115.5&(-116.6) \\
${^1}$TS\_POH-HPO  & -151.3 & -166.0 & -170.0\\
${^1}$HPO          & -432.9 & -444.1 & -440.5\\
${^1}$TS\_H-PO     &  -59.8 &  -82.0 & -97.6 &(-98.3) \\
${^2}$PO + ${^2}$H & -144.8 & -151.5 & -166.5 \\
${^1}$PH + O(${^1}D$) & 343.5 & --- & ---\\
\enddata
\tablecomments{${^a}$rev-DSD-PBEP86-D3(BJ)/jun-cc-pV(T+d)Z; see text.
${^b}$CCSD(T)/CBS+CV+DBOC+Rel+fT+fQ; CCSD(T)/ CBS+CV+DBOC+Rel+fT+fQ+fP results given within parentheses; see text.
${^c}$ CCSD(T)/CBS+CV+DBOC+Rel+fT+fQ + ZPE\textsubscript{anh}; CCSD(T)/CBS+CV+DBOC+Rel+fT+fQ+fP + ZPE\textsubscript{anh} results given within parentheses; see text.
$^d$ Referred to $^4$P + $^2$OH.}
\end{deluxetable}

From the pre-reactive complexes, ${^3}$POH and ${^1}$POH, the PESs bifurcates into two pathways. The first exit channel leads to ${^2}$PO + ${^2}$H through the transition structures ${^3}$TS$\_$PO-H and ${^1}$TS$\_$PO-H, which correspond to the dissociation of the O-H bond. The energy difference between these saddle points is small (18.1 kJ/mol). However, the energy barrier ruling the ${^3}$POH dissociation is 227.3 kJ/mol, which is reduced to 181.0 kJ/mol for ${^1}$POH. This difference in the barriers is mainly due to the fact that ${^3}$POH lies 64.4 kJ/mol below ${^1}$POH. Although the barrier for the triplet PES is higher than that for the singlet, this can be easily overcome because the system accumulates the energy of the reactants, which cannot be dissipated by third-body collisions due to the very low pressures of the ISM.

The second pathway proceeds with the isomerization of the pre-reactive complexes to ${^3}$HPO and ${^1}$HPO for the triplet and singlet PES, respectively. The energy difference between the corresponding transition states, ${^3}$TS$\_$POH-HPO and ${^1}$TS$\_$POH-HPO, is again small (16.7 kJ/mol), but the energy barriers differ considerably because of the relative stability of ${^3}$POH and ${^1}$POH. These are 174.2 kJ/mol and 126.5 kJ/mol for the triplet and singlet PES, respectively. The PESs for the hydrogen migration are very different for the two spin states: the isomerization of ${^1}$POH to ${^1}$HPO is exothermic (-144.0 kJ/mol), whereas from ${^3}$POH to ${^3}$HPO the process is endothermic (91.1 kJ/mol). These findings are in good agreement with those reported in \citet{FRANCISCO2003303}.
The last part of the mechanism is the dissociation of the P-H bond in the intermediates ${^3}$HPO and ${^1}$HPO, which is the second exit channel. While in the triplet PES the process is barrierless, in the singlet PES, the dissociation of the P-H in ${^1}$HPO goes through a high energy barrier (342.9 kJ/mol), ruled by the ${^1}$TS$\_$H-PO transition state. We have also considered the dissociation of ${^3}$HPO and ${^1}$HPO to ${^1}$PH + O(${^3}P$) and ${^1}$PH + O(${^1}D$), respectively. However, these dissociation paths are endothermic processes (revDSD energies are given in Table~\ref{tab1}), which are thus not viable in the extreme conditions of the ISM. Therefore, they have not been further investigated. 

Overall, we can conclude that, from a thermochemical point of view, the formation of ${^2}$PO from the reaction between ${^2}$OH and P(${^4}S$) is feasible, the rate-determining step of the whole process being the barrierless association between the reactants. On the other hand, the formation of ${^2}$PO on the singlet PES is unlikely, since the rate-determining step is the formation of ${^1}$POH, which is ruled by a non-negligible energy barrier. The energy profile depicted in Figure~\ref{fig1} shows some regions where the two PESs could cross through minimum energy crossing points (MECP). These are the first exit channel and the isomerizations of ${^3}$POH and ${^1}$POH. However, as mentioned above, the most relevant parts of the whole processes are the association reactions to yield the intermediates ${^3}$POH and ${^1}$POH. In these first steps, the potential energies of the two PESs are wsell separated and they cannot cross. In addition, the energy of the reactants is very high in comparison with the rest of the stationary points. Therefore, crossings in the pathways, if any, are expected not to influence the kinetics of the reaction.

The relative energies of the paths discussed above are collected in Table~\ref{tab1}, where the results for revDSD and the HEAT-like composite scheme are compared. Such a comparison shows average deviations of 10-15 kJ/mol, thus suggesting that revDSD is suitable for semi-quantitative investigations. The HEAT-like data augmented for ZPE corrections evaluated at an anharmonic level are also provided. A note on the pentuples correction is also deserved. Their contribution to relative energies is smaller than 1 kJ/mol even for the species showing the largest multi-reference character, thus supporting the conclusion that the full treatment of quadruple excitations is already able to incorporate non-dynamical correlation in the energy determination.

\subsection{Reaction of P($^4$S) and P($^2$P) with water (H$_2$O)}

\begin{figure}[t!]
  \includegraphics[width=1.1\textwidth]{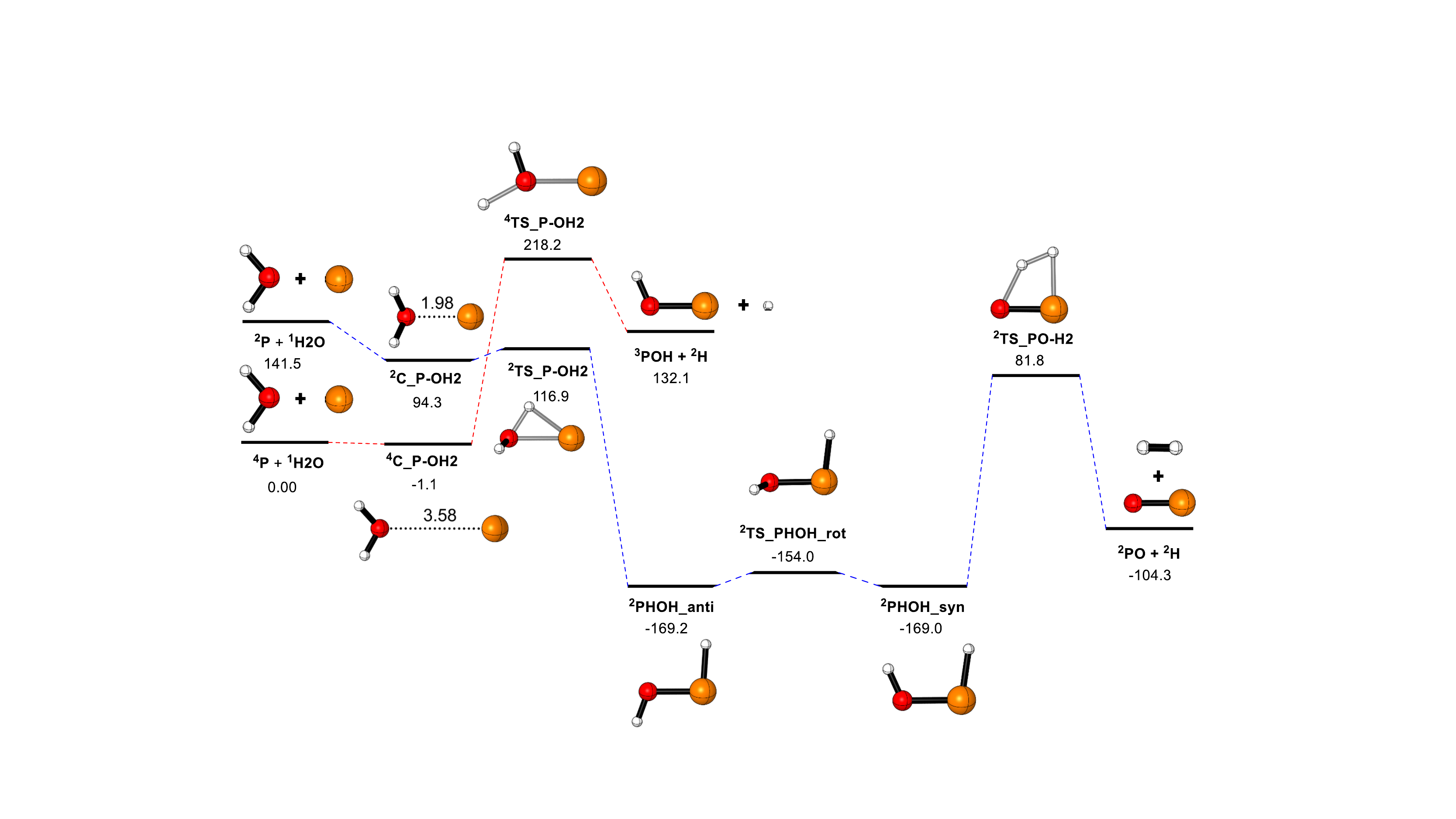}
  \caption{Doublet (blue profile) and quartet (red profile) potential energy surfaces (ZPE-corrected energies) for the formation of PO from H$_{2}$O +P.}
  \label{fig2}
\end{figure}

The reactivity of atomic phosphorus with H$_{2}$O is very different with respect to that with OH. First of all, the high-spin (quartet) PES corresponding to the reaction between H$_{2}$O and P(${^4}S$) is not expected to proceed toward the desired product, i.e. PO, under the conditions of the ISM. In fact, after the barrierless formation of a van der Waals pre-reactive complex (${^4}$C$\_$P-OH2), whose relative energy is only $\sim$1 kJ/mol below the reactants, the process toward the triplet ${^3}$POH and atomic hydrogen (${^2}$H) goes through a transition state ${^4}$TS$\_$P-OH2 (with a barrier of 213.3 kJ/mol). According to the discussion of the previous section (see also Figure~\ref{fig1}), two possible reaction pathways are available from ${^3}$POH to ${^2}$PO + ${^2}$H. However, both of them are ruled by non-negligible energy barriers. The path from H$_{2}$O + P(${^4}S$) to ${^3}$POH + H is depicted in Figure~\ref{fig2} (red profile).

From a chemical point of view, the reaction between H$_2$O and P($^2P$) is instead expected to occur in the ISM. However, it requires the presence of atomic phosphorus in an excited electronic state, whose abundance should be negligible in the ISM with respect to that of the ground state. The reaction mechanism starts with the formation of the pre-reactive complex ${^2}$C$\_$P-OH2 (47.2 kJ/mol below the reactants, 94.0 kJ/mol above the reactants in the quartet state), which evolves to $^2$PHOH-anti by overcoming the transition state ${^2}$TS$\_$P-OH2 (with a barrier of 22.6 kJ/mol), which lies 24.6 kJ/mol below the reactants of the doublet PES. Then, ${^2}$PHOH-anti can isomerize to ${^2}$PHOH-syn (nearly isoenergetic with anti once ZPE is incorporated) by overcoming a small torsional barrier (18.3 kJ/mol, with ${^2}$TS$\_$PHOH$\_$rot lying 12.5 kJ/mol below the doublet reactants). The exit channel is the loss of molecular hydrogen from ${^2}$PHOH-syn to give ${^2}$PO + H$_{2}$ and it is characterized by a relatively high barrier (${^2}$TS$\_$PO-H2) of about 251 kJ/mol.  The path from H$_{2}$O + P(${^2}S$) to PO + H$_2$ is depicted in Figure~\ref{fig2} (blue profile). As noted for the reaction of ${^2}$OH + P(${^4}S$), the rate-determining step of this mechanism is the barrierless association between the reactants. In passing we noted that the dissociation of either ${^2}$PHOH-syn or ${^2}$PHOH-anti to ${^1}$PH + ${^2}$OH has been investigated. As noted for the P + OH reaction, the formation of PH is an endothermic process (revDSD energies are available in Table~\ref{tab2}). Therefore, it has not been further considered.

In analogy with the P + OH reaction, we collect and compare in Table~\ref{tab2} the relative revDSD and HEAT-like energies of the paths discussed above. The comparison shows a better agreement between revDSD and HEAT-like results with respect to the P + OH reactions, with an average deviation of the order of a few kJ/mol. The better agreement is probably due to the fact that, for the reaction under consideration, the T$_1$ diagnostics did not point out any multireference character. In Table~\ref{tab2}, the HEAT-like values corrected for the ZPE contribution evaluated at an anharmonic level are also listed.

\begin{deluxetable}{cccc}
\label{tab2}
%\tablenum{2}
\tablecaption{Relative energies (in kJ/mol) for the P(${^4}S$)/P(${^2}P$) + H$_2$O reaction.}
\tablewidth{0pt}
\tablehead{
%\colhead{Stationary points} & \multicolumn{2}{c}{Energy (kJ/mol)} & \colhead{Energy + ZPEanh corrected (kJ/mol)}}\\
 & revDSD${^a}$ & HEAT-like${^b}$ & HEAT-like + ZPE\textsubscript{anh}${^c}$}
 
\startdata
\hline
Quartet PES\\
\cline{1-1}
${^4}$P + H${_2}$O  &   0.0 & 0.0   & 0.0 \\
${^4}$C\_P-OH2      &  -2.0 & -1.7  & -1.1\\
${^4}$TS\_P-OH2     & 231.4 & 237.7 & 218.2\\
${^3}$POH + ${^2}$H & 149.8 & 154.8 & 132.1\\
\hline 
Doublet PES${^d}$\\
\cline{1-1}
${^2}$P + H${_2}$O  &  173.7 &  141.5 &  141.5\\
${^2}$C\_P-OH2      &   72.7 &   85.8 &   94.3\\
${^4}$TS\_P-OH2     &  125.8 &  123.6 &  116.9\\
${^2}$PHOH\_anti    & -166.8 & -168.7 & -169.2\\
${^2}$TS\_PHOH\_rot & -152.5 & -150.9 & -154.0\\
${^2}$PHOH\_syn     & -170.9 & -169.9 & -169.0\\
${^2}$TS\_PO-H2     &   95.1 &   97.0 &   81.8\\
${^2}$PO + H${_2}$  &  -81.1 &  -82.2 & -104.3\\
${^1}$PH + ${^2}$OH & 206.5 & --- & ---\\
\enddata
\tablecomments{${^a}$rev-DSD-PBEP86-D3(BJ)/jun-cc-pV(T+d)Z; see text. \\
${^b}$CCSD(T)/CBS+CV+DBOC+Rel+fT+fQ; see text. \\
${^c}$ CCSD(T)/CBS+CV+DBOC+Rel+fT+fQ + ZPE\textsubscript{anh}; see text. \\
${^d}$ Referred to ${^4}$P + H$_2$O.}
\end{deluxetable}

\subsection{Kinetics}
\label{kinetics}

\begin{figure}[t!]
  \includegraphics[width=1.1\textwidth]{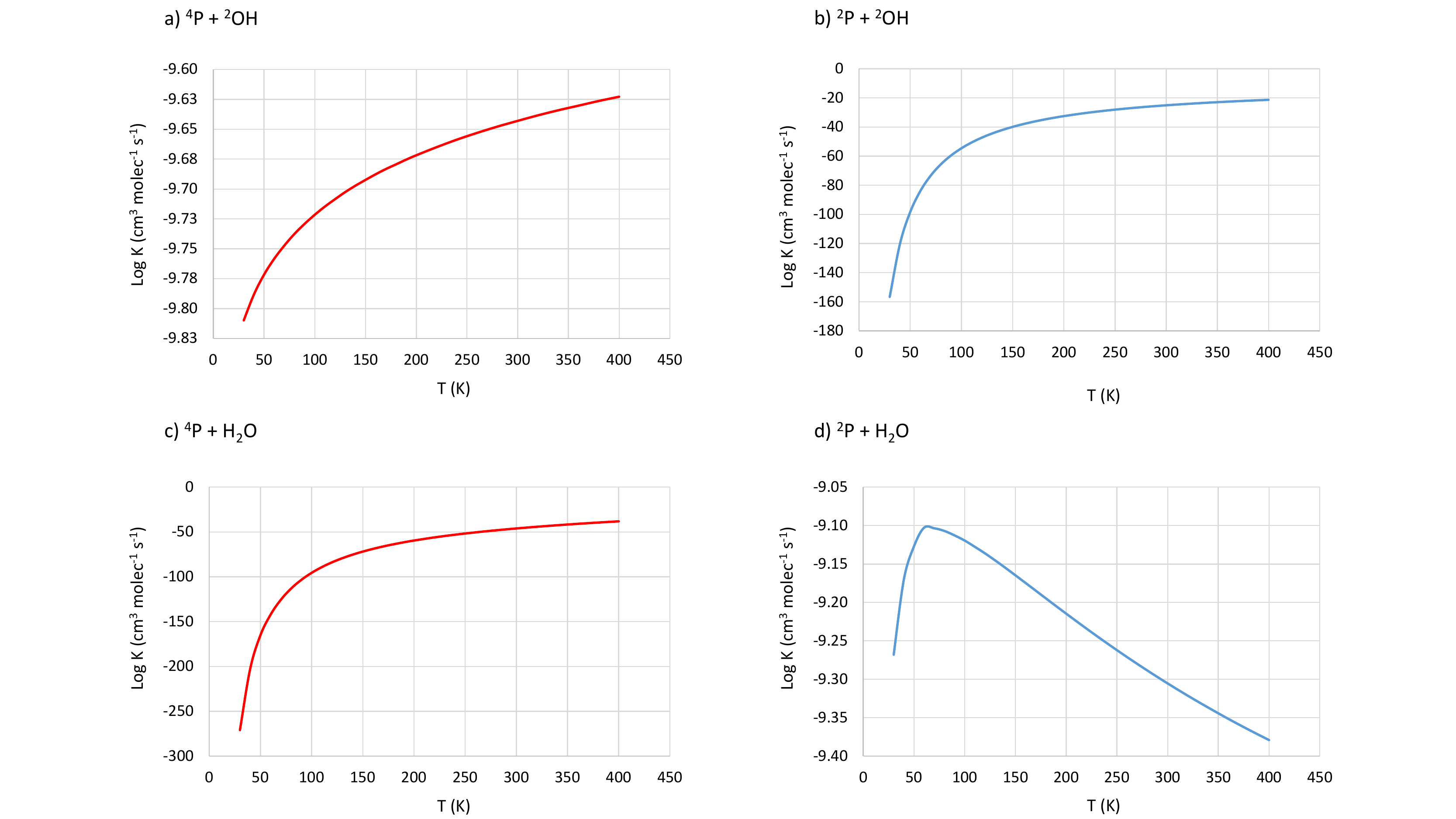}
  \caption{Plots of global rate coefficients (log scale) for the P+OH and P+H$_2$O reactions in the 30-400 K.}
  \label{fig3}
\end{figure}

While thermochemistry provides hints on the viability of the studied reactions, to quantitatively understand whether they reasonably proceed toward the desired products, global rate constants need to be calculated. As described in the methodology section, we have resolved the one-dimensional master equation in the 30-400 K temperature range and at pressures of 10$^{-7}$ atm. Figure~\ref{fig3} shows the plots of the bimolecular rate constants as a function of the temperature for the four reaction channels studied. The a) and b) panels refer to the reactions of ${^4}$P and ${^2}$P with ${^2}$OH respectively, whereas the c) and d) panels refer to the reactions of ${^4}$P and ${^2}$P with H${_2}$O, respectively. In addition, for the four reactions, Tables~\ref{tab5} (Appendix) and~\ref{tab3} report the global rate constants in the temperature range considered and the resulting parameters of the Arrhenius-Kooij equation, respectively. 

As mentioned in Section 3.1, the rate-determining step for the reaction of ${^2}$OH with both ${^2}$P and ${^4}$P is the formation of the POH adduct, with all the other reaction steps being characterized by submerged barriers. 
The entrance channel of the singlet PES shows a transition state (${^1}$TS$\_$P-OH) that lies at 85.0 kJ/mol above the reactants in the doublet state. This makes the reaction very slow, as demonstrated by the extremely small global rate coefficients, which lie between 0.0 (at 30 K) and 5.39$\times$10$^{-22}$ (at 400 K) cm$^3$ molecule$^{-1}$ s$^{-1}$. The temperature dependence is well reproduced by the Arrhenius-Kooij model (see Eq.~\ref{AK}), as can be seen in Figure~\ref{fig3}, with fitting parameters  \(\alpha\)=2.97$\times$10$^{-11}$ cm${^3}$ molecule$^{-1}$ s$^{-1}$, \(\beta\)=1.11 and \(\gamma\)=1.00$\times$10$^{4}$ K. We note a strong temperature dependence from 30 K to 100 K (the curve being very steep), while above the latter temperature the variation of the rate coefficient is less evident. A similar trend is also observed for the triplet PES, even if --in such a case-- the temperature dependence is smoother in the entire range considered. Since the entrance channel of the ${^4}$P + OH reaction is barrierless, the global rate coefficients are greater and range between 1.55$\times$10$^{-10}$ and 2.38$\times$10$^{-10}$ for temperatures between 30 and 400 K. These global rate coefficients indicate that this pathway to PO is very efficient at the temperatures typical of the ISM The derived \(\alpha\), \(\beta\) and \(\gamma\) fitting parameters  are 2.28$\times$10$^{-10}$ cm${^3}$ molecule$^{-1}$ s$^{-1}$, 0.16 and 0.37 K, respectively. Finally, from the inspection of  Figure~\ref{fig1}, we note that the backward reactions are not feasible from any intermediate because the reactants are much higher in energy than any other minimum.

%EXAMPLE for FIGURES
%\begin{figure}
%\begin{center}
% \includegraphics[scale=0.27]{figure/XX}
% \end{center}
% \caption{caption .....}
% \label{fig:XX}
%\end{figure}

Moving to the reaction with water, we note that when considering the high spin atomic phosphorus (${^4}$P), the formation of the intermediate ${^3}$POH + H is not feasible at low temperature because of the high energy of the transition state ${^4}$TS$\_$PO-H2 (see section 3.2). This reaction also shows an Arrhenius behavior similar to that of ${^2}$P+OH, with the rate coefficients increasing by increasing the temperature, although the rate coefficients remain extremely small. Indeed, they range from 0.0 at 30 K to 6.31$\times$10$^{-39}$ cm$^3$ molecule$^{-1}$ s$^{-1}$ at 400 K (see Table~\ref{tab3}). These low rate constants are well reproduced by the \(\alpha\), \(\beta\) and \(\gamma\) parameters of the Arrhenius-Kooij expression, these being 3.00$\times$10$^{-25}$ cm${^3}$ molecule$^{-1}$ s$^{-1}$, 14.87 and 1.48$\times$10$^{4}$ K. Contrary to the reaction with OH, the association between the first metastable excited state of atomic phosphorus (${^2}$P) and water presents a non-Arrhenius behavior from 70 K to higher temperatures (Figure ~\ref{fig3}d). This change in the temperature dependence is due to the fact that the pre-reactive complex and following transition structure are very close in energy, and thus the actual temperature affects the easiness in overcoming the corresponding barrier. However, the rate-determining step of the whole process is the first barrierless association, and therefore this reaction channel is open at all temperatures here considered, with global rate coefficients ranging from 5.39$\times$10$^{-10}$ to  4.18$\times$10$^{-10}$ cm$^3$ molecule$^{-1}$ s$^{-1}$. The fitting parameters that describe this temperature dependence rate constants are \(\alpha\)=5.88$\times$10$^{-10}$ cm${^3}$ molecule$^{-1}$ s$^{-1}$, \(\beta\)=0.71 and \(\gamma\)=51.47 K.
For the reaction of water with ${^2}$P, the reactants energy is well above all products and intermediates, and thus the reverse reaction can be considered not feasible. Indeed, the backward mechanism shows high energy transition states that cannot be overcome under typical ISM conditions.  

\begin{deluxetable*}{ccccc}
%\tablenum{1}
\label{tab3}
\tablecaption{Arrhenius-Kooij parameters for the P+OH and P+H$_2$O reactions.}
\tablewidth{0pt}
\tablehead{\colhead{Fitting parameters} & \colhead{${^4}$P + ${^2}$OH} & \colhead{${^2}$P + ${^2}$OH} & \colhead{${^4}$P + H${_2}$O} & \colhead{${^2}$P + H${_2}$O}} 
\startdata
$\alpha$ / cm${^3}$ molecule$^{-1}$ s$^{-1}$ & 2.28$\times$10$^{-10}$  & 2.97$\times$10$^{-11}$ & 3.00$\times$10$^{-25}$ & 5.88$\times$10$^{-10}$ \\
$\beta$   & 0.16      & 1.11     & 14.87    & -0.71\\
$\gamma$ / K & 0.37      & 1.00$\times$10$^4$  & 1.48$\times$10$^4$  & 51.47\\
\enddata
\end{deluxetable*}

\section{Discussion}
\label{sec:discussion}

\subsection{Comparison of the rate constants of the P + OH and N + OH reactions}

As mentioned in Section~\ref{sec:intro}, the abundance of PO in the ISM is higher than that of PN in regions affected by shocks \citep[either associated with molecular outflows or large-scale shocks in molecular clouds in the Galactic Center; see e.g.][]{lefloch16,bergner19,victor16,victor2018}. The observed abundance of PO with respect to PN is a useful indicator because it provides information about the length of the pre-stellar collapse phase \citep[see][]{Aota,lefloch16}. However, the lack of well-validated kinetic data for the formation of PO hampers our understanding of the reason why PO consistently appears to be more abundant than PN in the ISM. Since the reaction kinetic information on P + OH $\rightarrow$ PO + H was missing prior to this work, in order to incorporate it in the modelling of phosphorus chemistry in the ISM, \citet{Izaskun18} assumed a rate constant for the association of P and OH similar to that of the well-studied N(${^4}S$) + OH(${^2}\Pi$) reaction \citep[see][and references therein]{daranlot11}. To check the validity of this assumption, we here compare the rate constant of the reaction N(${^4}S$) + OH(${^2}\Pi$) with that calculated in Section~\ref{kinetics} for the P + OH reactive system. 

\begin{figure}[b!]
  \includegraphics[width=1.05\textwidth]{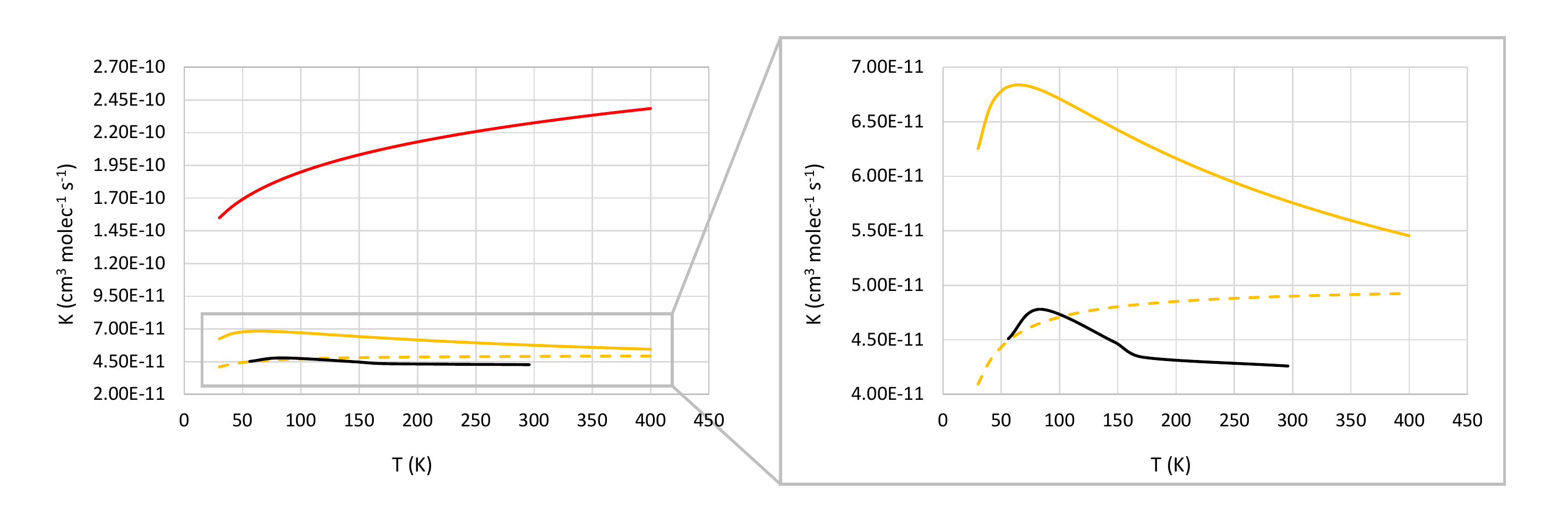}
  \caption{Plots of the temperature dependence of the global rate coefficients for the P(${^4}S$) + OH(${^2}\Pi$) reaction (red line) compared with those, obtained from the Arrhenius-Kooij fitting parameters taken from UMIST (yellow solid line) and from KIDA (yellow dotted line), for the N(${^4}S$) + OH(${^2}\Pi$) reaction are shown. Experimental rate coefficients for the latter reaction \citep{daranlot11} are also plotted (black line).}
  \label{fig4}
\end{figure}

In Figure~\ref{fig4}, we show the temperature dependence of the global rate constants for the reaction P(${^4}S$) + OH(${^2}\Pi$) $\rightarrow$ ${^2}$PO + ${^2}$H (red line; this work), and those available in the literature for the reaction N(${^4}S$) + OH(${^2}\Pi$) $\rightarrow$ NO(${^2}\Pi$) + ${^2}$H. For the latter reaction, we compare the Arrhenius-Kooij fitting parameters provided by the UMIST (University of Manchester Institute of Science and Technology \citep{umist}; solid yellow line) and KIDA  (KInetic Database for Astrochemistry \citep{kida}; dotted yellow line) databases with the most recent experimental values for this reaction (\citet{daranlot11}; black line). While the order of magnitude of the KIDA results is closer than the UMIST counterparts to the experimental values, the temperature dependence of the rate constant is better described by the UMIST data, which correctly reproduce the non-Arrhenius behavior observed in the experiment (in the $\sim$80-300$\,$K range). On the contrary, the global rate constant of the P + OH reaction shows an Arrhenius-like temperature dependence (see red line in Figure~\ref{fig4}, and Section~\ref{kinetics}), and it is about one order of magnitude greater than that of the reaction between N and OH.

The explanation of the behaviors and results discussed above lies in the fact that P(${^4}S$) and N(${^4}S$) show very similar reaction mechanisms when they react with OH(${^2}\Pi$) \citep{li11}, but the exit channels present significant differences. While the exit channel yielding PO + H from HPO is barrierless (see Figure~\ref{fig1}), the dissociation of HNO leading to NO(${^2}\Pi$) + ${^2}$H shows a transition state (see Figure 5 in \citet{li11}), which is likely the cause of the non-Arrhenius behaviour of the reaction N(${^4}S$) + OH(${^2}\Pi$) at temperatures above 80 K. Furthermore, the presence of such a barrier in the exit channel for the N + OH reaction might also explain the greater global rate constant when OH reacts with P. We can thus conclude that the N(${^4}S$) + OH(${^2}\Pi$) reaction is not a good model for the P(${^4}S$) + OH(${^2}\Pi$) reaction studied in this work, because the rate coefficients of the two reactions great differ in the absolute value as well as in the temperature dependence.

\subsection{Astrophysical implications for the P + H$_2$O reaction}

Water is known to be ubiquitous in the ISM \citep[][]{vandishoeck21}, and therefore it plays an important role in the oxidization of chemical species in the Universe. In the case of phosphorus, the oxidation reaction of P in its ground state (${^4}S$) with water to yield ${^2}$PO does not take place in the ISM given its extremely low rate constant at all temperatures (see Table~\ref{tab5} in the Appendix as well as Figures~\ref{fig2} and~\ref{fig3}c). However, the reaction of H${_2}$O with the first excited state P(${^2}P$) is highly efficient (see Figure~\ref{fig3}d), although most of the atomic phosphorus in the ISM should be in its ground state. 

Water is also expected to be abundant in shocked regions associated with molecular outflows \citep{vandishoeck21,nisini13,dionatos20,suutarinen14,vandishoeck13}. However, even in regions with high abundances of water, the P(${^4}S$) + H${_2}$O reaction is inefficient, the rate constant being $\sim$10$^{-50}$~cm$^{3}$~molecule$^{-1}$~s$^{-1}$ at temperatures in the 200-300$\,$K range. This means that the formation of PO in the ISM through this chemical route is negligible. Water, however, can be either chemically destroyed at high temperatures \citep{Viti11} or UV photon-dissociated by the radiative precursor of J-type shocks \citep{tappe08}, both processes yielding OH. Therefore, one can expect that the formation of PO through the P + OH reaction becomes dominant in regions with large amounts of water, this latter acting as a precursor of OH. In the next section, this hypothesis will be investigated in further detail.

\subsection{Chemical modelling in a shocked region}

In this Section, we address the modelling of the phosphorus chemistry of an outflow shocked region with and without considering the PO formation routes explored in this work. As reference model, we will consider the C-type shock model that best reproduces the molecular abundances measured toward the B1 shocked region in the L1157 molecular outflow \citep[see][]{Viti11,holdship16}.

\begin{deluxetable*}{lccccr}
%\tablenum{3}
\label{tab4}
\tablecaption{Model parameters assumed for the L1157-B1 C-type shock.}
\tablewidth{0pt}
\tablehead{Parameters } 
\startdata
%\hline
n(H)        &&&&& 2$\times$10$^{5}$~cm$^{-3}$ \\
v$_s$       &&&&& 40$\,$km$\,$s$^{-1}$ \\
T$_{n,max}$ &&&&& 4000$\,$K \\
B$_0$       &&&&& 450~$\mu$G \\
t$_{sat}$   &&&&& 4.6$\,$yrs \\
\enddata
\end{deluxetable*}

For the modelling of the phosphorus chemistry, we use the UCLCHEM chemical code \citep{uclchem} and the phosphorus chemical network built by \citet{Izaskun18}. This chemical network is here updated by incorporating the rate constants of the P(${^4}S$) + OH(${^2}\Pi$) $\rightarrow$ ${^2}$PO + ${^2}$H and P(${^4}S$) + H$_2$O $\rightarrow$ ${^2}$PO + H${_2}$  processes (see Table~\ref{tab4}). UCLCHEM is run in three phases: Phase 0 simulates the chemistry of a diffuse molecular cloud for 10$^6 \,$yrs assuming a n(H) density of 10$^3 \,$cm$^{-3}$ and a temperature of 20$\,$K. In Phase 1, the cloud undergoes free-fall collapse at a constant temperature of 10 K until the final density of 2$\times$10$^5 \,$cm$^{-3}$ is reached. Finally, Phase 2 simulates the physical processes associated with the passage of a magneto-hydrodynamic (C-type) shock, which are commonly found in molecular outflows. For the physical structure of the C-type shock, we employ the same parameters used by \citet{Izaskun182}. Table~\ref{tab4} reports the input parameters of the L1157-B1 shock model, where n(H) refers to the initial volume density of the gas (2$\times$10$^{5} \,$cm$^{-3}$), v$_s$ is the shock speed (40$\,$km$\,$s$^{-1}$), T$_{n,max}$ is the maximum temperature achieved by the neutral gas in the post-shock region (4000$\,$K), B$_0$ is the magnetic field strength (450~$\mu$G) and t$_{sat}$ is the time at which the majority of the ice mantles of dust grains are released back into the gas phase due to sputtering (of 4.6$\,$yrs for a n(H) density of 2$\times$10$^{5} \,$cm$^{-3}$; see \citet{Izaskun18}). For our shock model, we also consider a long-lived collapse of 6$\times$10$^6 \,$yrs, i.e. once the final density is reached, it remains constant for about 7$\times$10$^5 \,$yrs longer. In addition, we assume a standard UV interstellar radiation field (G$_0$=1 in Draine units; see \citet{viti99}) and a standard cosmic-ray ionization rate ($\zeta$=1.3$\times$10$^{-17} \,$s$^{-1}$). The assumed initial abundance of P is 2.57$\times$10$^{-9}$, as inferred by \citet{Aota} and \citet{lefloch16} for the L1157-B1 shock.\footnote{Note that, in order to reproduce the measured abundances of PN and PO toward L1157-B1, atomic phosphorus needs to be depleted by a factor of 100 with respect to its solar value; otherwise, the abundances of these P-bearing species are overestimated \citep{Aota,Lefloch}}
For further details on the modelling, the reader is referred to \citet{Izaskun182}.

\begin{figure}[t!]
\begin{center}
  \includegraphics[angle=0,width=0.8\textwidth]{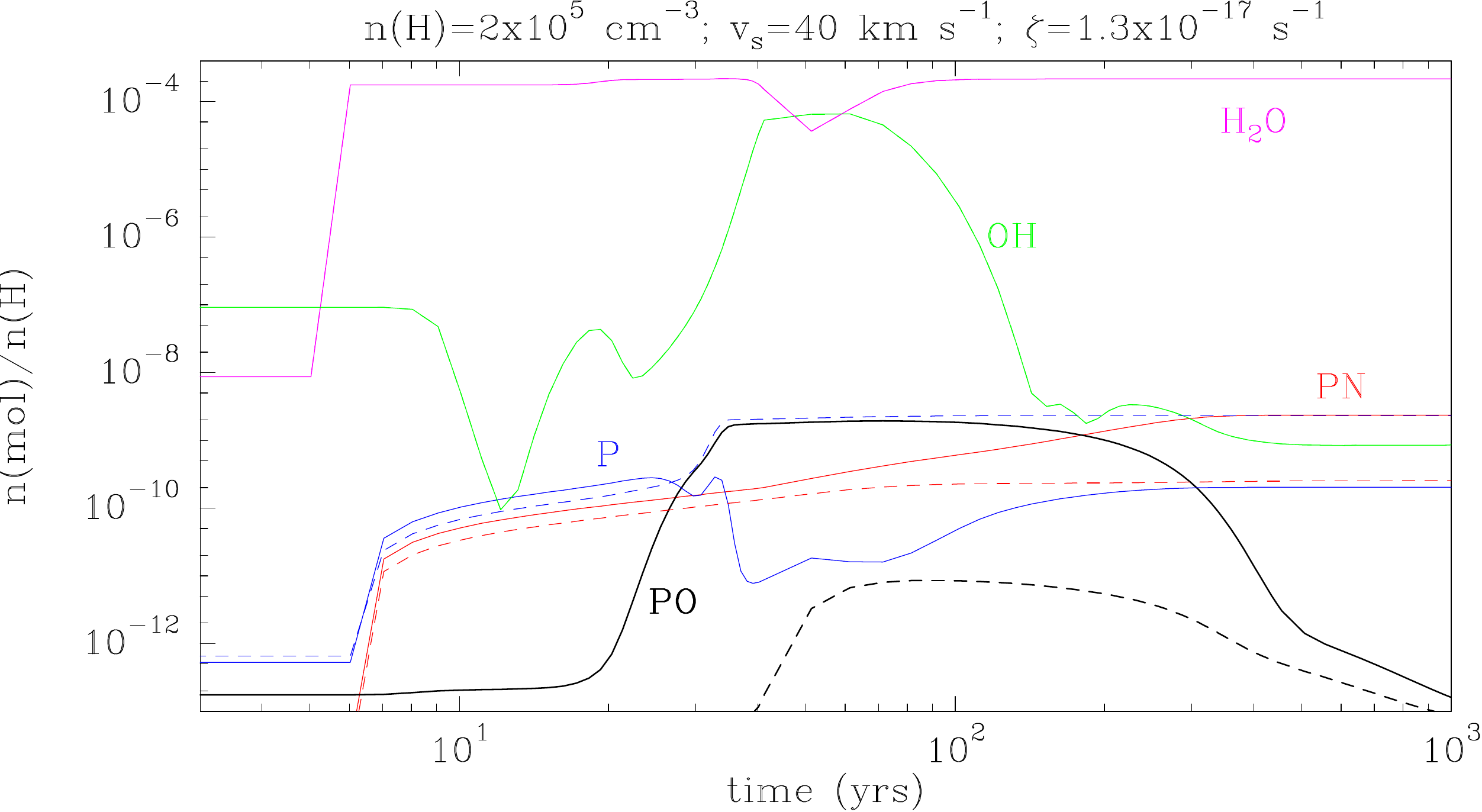}
  \caption{Abundances of P, PO, PN, OH and H$_2$O predicted for Phase 2 of our model and assuming the shock parameters collected in Table~\ref{tab4}. These shock parameters are those that best reproduce the molecular abundances measured in the L1157-B1 shocked region \citep{Viti11,holdship16}. Solid lines refer to the results from the model {\it with} the new PO rate constants calculated in this work for the P + OH $\rightarrow$ PO + H and P + H$_2$O $\rightarrow$ PO + H$_2$ reactions. Dashed lines show the results for the model {\it without} the new rate constants for the formation of PO. The abundances of OH and H$_2$O are the same in both models.}
  \label{fig5}
\end{center}
\end{figure}

In Figure~\ref{fig5}, we present the results of the models of the phosphorus chemistry for the C-type shock parameters shown in Table~\ref{tab4}. The abundances of the model {\it with} the new rate constants for the formation of PO, calculated in this work, are shown with solid lines, while the results from the model {\it without} these new rate constants are presented in dashed lines. The abundance of H$_2$O, P and PN increase at the beginning of the shock (at about time$\sim$5$\,$yrs) due to the release of the icy mantles by the sputtering of dust grains (i.e. when time$\geq$t$_{sat}$; see \citet{Izaskun18}). The enhancement of P is tightly linked to the release of PH$_3$ from dust grains, since the latter is rapidly converted into PH$_2$, PH and then P through the endothermic reactions PH$_3$ + H $\rightarrow$ PH$_2$ + H$_2$, PH$_2$ + H $\rightarrow$ PH + H$_2$ and PH + H $\rightarrow$ P + H$_2$ (these reactions have energy barriers from 318 to 735$\,$K; see \citet{charnley94}). 

The inclusion of the P + OH $\rightarrow$ PO + H and P + H$_2$O $\rightarrow$ PO + H$_2$ reactions in the chemical network does not modify noticeably the abundance of P and PN in the collapse phase, as shown by the similar abundances of these molecules at the beginning of the shock for both models (Figure~\ref{fig5}). However, the abundance of PO is significantly enhanced not only during the collapse phase but, more importantly, in the post-shocked gas owing to the new formation route P + OH $\rightarrow$ PO + H. Figure~\ref{fig5} indeed shows that the PO abundance is enhanced by three orders of magnitude in the post-shocked gas (from a few 10$^{-12}$ to a few 10$^{-9}$), thereby reaching values consistent with those measured in L1157-B1 \citep[of $\sim$2.5$\times$10$^{-9}$; see][]{lefloch16}. The enhancement of PO in the post-shocked gas is also possible thanks to the slight drop in the abundance of H$_2$O that increases the abundance of OH by several orders of magnitude when the temperature of the shocked gas reaches its maximum value \citep[at about 50$\,$yrs after the passage of the shock; see also][]{Izaskun182}. Note that the P + OH $\rightarrow$ PO + H reaction also seems to be the main mechanism responsible for the destruction of atomic phosphorus in the post-shocked gas (see blue lines in Figure~\ref{fig5}). The predicted PO/PN ratios range between 3 and 9 in the post-shock region where PO is more abundant than PN (see Figure~\ref{fig5}), which is also consistent with the observations \citep[][]{lefloch16,victor16,victor2018,Victor2020,bergner19,bernal21}. 

\section{Conclusions}  
\label{sec:conclusions}

In this paper we have studied the formation route of the phosphorus monoxide (${^2}$PO) from the reaction of the oxygenated species H${_2}$O and OH(${^2}\Pi$) with both the ground (${^4}S$) and the first metastable excited state (${^2}P$) of atomic phosphorus (P). Our theoretical results indicate that the formation of PO in the gas phase from the P(${^4}S$) + OH reaction is highly efficient because the rate-determining step of the whole process is the barrierless association reaction of the reactants. On the contrary, the first metastable excited state P(${^2}P$) should not react with OH under the same conditions due to a high energy barrier in the entrance channel. For the reaction of P with H$_2$O, the reactivity is reversed: while the reaction with P(${^4}S$) does not take place under the typical conditions of the ISM, that with P(${^2}P$) is highly efficient, being ruled by a rate-determining barrierless association.
Overall, the reaction of atomic phosphorus in its ground state with OH is expected to be an important source of ${^2}$PO in the ISM because it enhances its abundance in shocked regions by several orders of magnitude. The modelled PO/PN ratios are indeed consistent with the values measured in the shocked gas associated with molecular outflows in star-forming regions.

\acknowledgments
This work has been supported by the Spanish State Research Agency (AEI) through project number MDM-2017-0737 Unidad de
Excelencia “María de Maeztu”—Centro de Astrobiología
(CSIC-INTA); by the MIUR ‘PRIN 2017’ (Grant Number 2017A4XRCA); by the Italian Space Agency (ASI; ‘Life in Space’ project, N. 2019-3-U.0); by AEI (PID2019-105552RB-C41). Partial support from the Spanish FEDER (ESP2017-86582-C4-1-R) is also acknowledged. Computational support was provided by the Supercomputer facilities of
LUSITANIA founded by Cénits and Computaex Foundation.
The SMART@SNS Laboratory (http://smart.sns.it) is acknowledged for providing high-performance computing facilities.





\clearpage

\appendix

%\section{Rate constants at different temperatures and PST coefficients}
%\label{appendix:kin}

\begin{deluxetable*}{ccccc}[h]
%\tablenum{3}
\label{tab5}
\tablecaption{Global rate constants (in cm${^3}$ molecule${^-1}$ s${^-1}$) evaluated at 1 x 10$^{^-7}$ atm.$^a$}
\tablewidth{0pt}
\tablehead{
T (K) & ${^4}$P + ${^2}$OH & ${^2}$P + ${^2}$OH & ${^4}$P + H${_2}$O & ${^2}$P + H${_2}$O} 
\startdata
\hline
30  & 1.55E-10 & 0.00  & 0.00  & 5.39E-10\\
40  & 1.63E-10  & 0.00 & 0.00 & 6.76E-10\\
50  & 1.69E-10 & 4.01E-99 & 0.00 & 7.46E-10\\
60 & 1.74E-10 & 1.56E-84 & 0.00  & 7.89E-10\\
70 & 1.79E-10 & 4.21E-74 & 0.00  & 7.88E-10\\
80 & 1.83E-10 & 2.95E-66 & 0.00 & 7.81E-10\\
90 & 1.87E-10 & 3.58E-60 & 0.00 & 7.71E-10\\
100 & 1.90E-10 & 2.71E-55 & 2.24E-96 & 7.59E-10\\
110 & 1.93E-10 & 2.72E-51 & 5.75E-90 & 7.45E-10\\
120 & 1.96E-10 & 5.93E-48 & 1.34E-84 & 7.31E-10\\ 
130 & 1.98E-10 & 4.00E-45 & 4.96E-80 & 7.15E-10\\
140 & 2.01E-10 & 1.07E-42 & 4.32E-76 & 7.00E-10\\
150 & 2.03E-10 & 1.37E-40 & 1.18E-72 & 6.84E-10\\
160 & 2.05E-10 & 9.57E-39 & 1.27E-69 & 6.69E-10\\
170 & 2.07E-10 & 4.08E-37 & 6.35E-67  & 6.53E-10\\
180 & 2.09E-10 & 1.15E-35 & 1.67E-64 & 6.38E-10\\
190 & 2.11E-10 & 2.29E-34 & 2.57E-62 & 6.24E-10\\
200 & 2.13E-10 & 3.38E-33 & 2.51E-60 & 6.10E-10\\
220 & 2.16E-10 & 3.57E-31 & 7.86E-57 & 5.83E-10\\
240 & 2.19E-10 & 1.75E-29  & 7.69E-54 & 5.58E-10\\
260 & 2.22E-10 & 4.74E-28 & 3.10E-51 & 5.35E-10\\
280 & 2.25E-10 & 8.04E-27 & 6.28E-49 & 5.14E-10\\
300 & 2.27E-10 & 9.41E-26 & 7.35E-47 & 4.95E-10\\
320 & 2.30E-10 & 8.13E-25 & 5.51E-45 & 4.77E-10\\
340 & 2.32E-10 & 5.47E-24 & 2.85E-43 & 4.60E-10\\
360 & 2.34E-10 & 2.98E-23 & 1.06E-41 & 4.45E-10\\
380 & 2.36E-10 & 1.37E-22 & 2.95E-40 & 4.31E-10\\
400 & 2.38E-10 & 5.39E-22 & 6.31E-39 & 4.18E-10\\
\enddata
\tablecomments{${^a}$ All values lower than 1.0E-100 have been set equal to zero.}
\end{deluxetable*}

\begin{deluxetable*}{ccccc}[b!]
%\tablenum{4}
\label{tab6}
\tablecaption{$C$ coefficients for PST calculations (in Hartree/bohr${^6}$) of the barrierless processes.}
\tablewidth{0pt}
\tablehead{Reaction &&&& Coefficient}
\startdata
\hline
${^4}$P + ${^2}$OH $\rightarrow$ ${^3}$POH        &&&& 112.479\\
${^3}$HPO $\rightarrow$ ${^2}$PO + ${^2}$H        &&&& 26.886\\
${^4}$P + H${_2}$O $\rightarrow$ ${^4}$C$\_$P-OH2 &&&& 112.067\\
${^2}$P + H${_2}$O $\rightarrow$ ${^2}$C$\_$P-OH2 &&&& 905.650\\
\enddata
\end{deluxetable*}

\clearpage

Cartesian coordinates (in angstrom) and harmonic frequencies (in cm$^{-1}$) of all stationary points 
\setlength{\parskip}{8mm}

${^2}$PO\\
P                  0.00000000    0.00000000    0.00000000\\
O                  0.00000000    0.00000000    1.48500000\\
freq(harm) = 1223.74    
\setlength{\parskip}{8mm}

${^2}$OH\\
 O                  0.00000000    0.00000000    0.00000000\\
 H                  0.00000000    0.00000000    0.97170000\\
freq(harm) = 3757.23  
\setlength{\parskip}{8mm}

${^1}$PH\\
 P                  0.00000000    0.00000000    0.08887700\\
 H                  0.00000000    0.00000000   -1.33315100\\
freq(harm) = 2407.47
\setlength{\parskip}{8mm}

${^1}$TS$\_$P-OH\\
 P                  3.46948464    0.00000000   -0.30171826\\
 O                  0.00000000    0.00000000    0.98010000\\
 H                  0.00000000    0.00000000    0.00000000\\
freq(harm) = 3638.141 270.532 225.949i
\setlength{\parskip}{8mm}

${^1}$POH\\
 P                  1.50696687    0.00000000    1.57166368\\
 O                  0.00000000    0.00000000    0.96660000\\
 H                  0.00000000    0.00000000    0.00000000\\
freq(harm) = 3760.799 1070.214 894.533
\setlength{\parskip}{8mm}

${^3}$POH\\
 P                  1.50076273    0.00000000    1.63486360\\
 O                  0.00000000    0.00000000    0.96420000\\
 H                  0.00000000    0.00000000    0.00000000\\
freq(harm) = 3823.637 933.103 827.148
\setlength{\parskip}{8mm}

${^1}$TS$\_$POH-HPO\\
 P                  0.00000000    0.00000000    1.52880000\\
 O                  1.18231570    0.00000000    0.41503159\\
 H                  0.00000000    0.00000000    0.00000000\\
freq(harm) = 2236.417 897.644 2010.099i
\setlength{\parskip}{8mm}

${^3}$TS$\_$POH-HPO\\
 P                  0.00000000    0.00000000    1.51110000\\
 O                  1.27629195    0.00000000    0.54335946\\
 H                  0.00000000    0.00000000    0.00000000\\
freq(harm) = 2134.077 944.020 1706.533i
\setlength{\parskip}{8mm}

${^1}$HPO\\
 P                  0.00000000    0.00000000    1.45680000\\
 O                  1.44309843    0.00000000    1.82608998\\
 H                  0.00000000    0.00000000    0.00000000\\
freq(harm) = 2169.554 1193.325 1008.357
\setlength{\parskip}{8mm}

${^3}$HPO\\
 P                  0.00000000    0.00000000    1.42860000\\
 O                  1.31934590    0.00000000    2.13424923\\
 H                  0.00000000    0.00000000    0.00000000\\
freq(harm) = 2248.498 1277.822 736.519
\setlength{\parskip}{8mm}

${^1}$TS$\_$H-PO\\
 P                  0.00000000    0.00000000    4.04900000\\
 O                  1.35434326    0.00000000    4.66632397\\
 H                  0.00000000    0.00000000    0.00000000\\
freq(harm) = 1199.271 88.235 512.679i
\setlength{\parskip}{8mm}

${^1}$TS$\_$PO-H\\
 P                  1.04552135    0.00000000    3.34359721\\
 O                  0.00000000    0.00000000    2.28130000\\
 H                  0.00000000    0.00000000    0.00000000\\
freq(harm) = 1233.134 363.761 394.336i
\setlength{\parskip}{8mm}

${^3}$TS$\_$PO-H\\
 P                  1.16667888    0.00000000    2.49535795\\
 O                  0.00000000    0.00000000    1.56920000\\
 H                  0.00000000    0.00000000    0.00000000\\
freq(harm) = 1286.444 484.534 2299.477i
\setlength{\parskip}{8mm}

H${_2}$O\\
O          0.00000        0.00000        0.11772\\
H          0.00000        0.75970       -0.47088\\
H          0.00000       -0.75970       -0.47088\\
freq(harm) = 3938.962 3824.073 1644.853 
\setlength{\parskip}{8mm}

${^2}$C$\_$P-OH2\\
P         -0.82055       -0.00001        0.00804\\
O          1.15974       -0.00002       -0.09907\\
H          1.51540       -0.78402        0.33604\\
H          1.51487        0.78431        0.33591\\
freq(harm) = 3885.769 3772.303 1610.048 654.769 539.814 383.317
\setlength{\parskip}{8mm}

${^2}$TS$\_$P-OH2\\
P         -0.75959       -0.02603       -0.00228\\
O          1.16713        0.01858       -0.11134\\
H          1.51114       -0.62345        0.52705\\
H          0.54567        0.86523        0.39787\\
freq(harm) = 3767.816 2110.363 940.039 822.651 637.448 1532.733i
\setlength{\parskip}{8mm}

${^2}$PHOH$\_$anti\\
P         -0.59530       -0.11409       -0.00000\\
O          1.04102        0.14268        0.00000\\
H          1.50919       -0.69911        0.00000\\
H         -0.90777        1.26899        0.00000\\
freq(harm) = 3843.236 2407.780 1129.724 920.899 826.633 433.988 
\setlength{\parskip}{8mm}

${^2}$TS$\_$PHOH$\_$rot\\
P          0.59726       -0.10260        0.01188\\
O         -1.04704        0.02831       -0.11618\\
H         -1.50635        0.02317        0.72984\\
H          0.92381        1.28946        0.02146\\
freq(harm) = 3842.659  2324.016 985.582 939.317 799.742 469.051i
\setlength{\parskip}{8mm}

${^2}$PHOH$\_$syn\\
P          0.60247       -0.08676       -0.00000\\
O         -1.04917       -0.09968        0.00001\\
H         -1.45144        0.77374        0.00002\\
H          0.80776        1.32520       -0.00001\\
freq(harm) = 3862.082 2347.653 1086.997 908.622 819.464 305.855 
\setlength{\parskip}{8mm}

${^2}$TS$\_$PO-H2\\
P         -0.53792       -0.13678       -0.00003\\
O          1.03330       -0.08189        0.00005\\
H          0.36012        1.17997        0.00003\\
H         -0.55775        1.52688       -0.00002\\
freq(harm) = 2074.884 1895.574 1063.123 958.833 853.231 1924.571i
\setlength{\parskip}{8mm}

H${_2}$\\
H          0.00000        0.00000        0.37110\\
H          0.00000        0.00000       -0.37110\\
freq(harm) = 2223.985
\setlength{\parskip}{8mm}

${^4}$C$\_$P-OH2\\
P         -1.47152       -0.00125        0.00437\\
O          2.10783       -0.00966       -0.06846\\
H          2.68106       -0.70795        0.25992\\
H          2.52906        0.80400        0.22228\\
freq(harm) = 3936.238 3821.003 1643.691 47.225 32.406 22.647
\setlength{\parskip}{8mm}

${^4}$TS$\_$P-OH2\\
P          0.78930       -0.01662       -0.00002\\
O         -1.03606       -0.00511        0.00003\\
H         -1.34418        0.92132        0.00004\\
H         -2.20681       -0.63113        0.00006\\
freq(harm) = 3639.759 1119.834 714.051 351.825 286.703 1854.174i
\setlength{\parskip}{8mm}

\clearpage

\bibliography{Astrochem}{}
\bibliographystyle{aasjournal}



\end{document}